\documentclass[useAMS,usenatbib,usegraphicx]{mn2e}

\def\eps@scaling{.95}
\def\epsscale#1{\gdef\eps@scaling{#1}}

\def\plotone#1{\centering \leavevmode
    \epsfxsize=\eps@scaling\columnwidth \epsfbox{#1}}

\title[2dFGRS: the luminosity function of cluster galaxies]
{The 2dF Galaxy Redshift Survey: the luminosity function of cluster galaxies}
\author[R. De Propris et~al. (the 2dFGRS Team)]{Roberto De Propris$^{1,2}$
Matthew Colless$^1$,
Simon P. Driver$^1$,
Warrick Couch$^2$,
\newauthor
John A. Peacock$^3$,
Ivan K. Baldry$^4$, 
Carlton M. Baugh$^5$,
Chris Collins$^6$,
\newauthor
Joss Bland-Hawthorn$^7$, 
Terry Bridges$^7$,
Russell Cannon$^7$, 
Shaun Cole$^5$,
Nicholas Cross$^4$, 
\newauthor
Gavin B. Dalton$^8$,
George Efstathiou$^{9}$,
Richard S. Ellis$^{10}$, 
Carlos S. Frenk$^5$, 
\newauthor
Karl Glazebrook$^4$,
Edward Hawkins$^{11}$,
Carole Jackson$^1$,
Ofer Lahav$^{8}$, 
Ian Lewis$^8$, 
\newauthor
Stuart Lumsden$^{12}$,
Steve Maddox$^{11}$, 
Darren S. Madgwick$^{9}$, 
Peder Norberg$^5$,
\newauthor
Will Percival$^{3}$,
Bruce Peterson$^1$, 
Will Sutherland$^7$, 
Keith Taylor$^{10}$\\
$^1$Research School of Astronomy \& Astrophysics, The Australian National 
University, Weston Creek, ACT 2611, Australia  \\
$^2$Department of Astrophysics and Optics, University of New South
Wales, Sydney, NSW 2052, Australia  \\
$^{3}$ Institute for Astronomy, University of Edinburgh, Royal Observatory, 
Edinburgh EH9 3HJ, United Kingdom  \\
$^4$Department of Physics \& Astronomy, Johns Hopkins University, 3400 
North Charles Street Baltimore, MD 21218, USA  \\
$^5$Department of Physics, University of Durham, Science Laboratories, 
South Road, Durham DH1 3LE, United Kingdom  \\
$^6$ Astrophysics Research Institute, Liverpool John Moores University, 
Twelve Quays House, Egerton Wharf, Birkenhead, L14 1LD, UK  \\
$^7$ Anglo-Australian Observatory, P.O. Box 296, Epping, NSW 2121, 
Australia  \\
$^8$Department of Physics, Keble Road, Oxford OX1 3RH, United Kingdom  \\
$^9$ Institute of Astronomy, University of Cambridge, Madingley Road, 
Cambridge CB3 0HA, United Kingdom  \\
$^{10}$ Department of Astronomy, California Institute of Technology, 
Pasadena, CA 91125, USA  \\
$^{11}$ School of Physics and Astronomy, University of Nottingham, 
University Park, Nottingham, NG7 2RD, United Kingdom  \\
$^{12}$ Department of Physics \& Astronomy, E C Stoner Building, 
Leeds LS2 9JT, UK
}

\begin{document}

\date{}

\pagerange{\pageref{firstpage}--\pageref{lastpage}} \pubyear{2003}

\maketitle

\label{firstpage}

\begin{abstract}
We have determined the composite luminosity function (LF) for galaxies
in 60 clusters from the 2dF Galaxy Redshift Survey. The LF spans the
range $-22.5<M_{b_{\rm J}}<-15$, and is well-fitted by a Schechter
function with ${M_{b_{\rm J}}}^{\!\!\!\!*}=-20.07\pm0.07$ and $\alpha=-1.28\pm0.03$
($H_0$=100\,km\,s$^{-1}$\,Mpc$^{-1}$, $\Omega_M$=0.3,
$\Omega_\Lambda$=0.7). It differs significantly from the field LF of
\cite{mad02}, having a characteristic magnitude that is approximately
0.3~mag brighter and a faint-end slope that is approximately 0.1
steeper. There is no evidence for variations in the LF across a wide
range of cluster properties: the LF is similar for clusters with high
and low velocity dispersions, for rich and poor clusters, for clusters
with different Bautz-Morgan types, and for clusters with and without
substructure. The core regions of clusters differ from the outer parts,
however, in having an excess of very bright galaxies. We also construct
the LFs for early (quiescent), intermediate and late (star-forming)
spectral types. We find that, as in the field, the LFs of earlier-type
galaxies have brighter characteristic magnitudes and shallower faint-end
slopes. However the LF of early-type galaxies in clusters is both
brighter and steeper than its field counterpart, although the LF of
late-type galaxies is very similar. The trend of faint-end slope with
spectral type is therefore much less pronounced in clusters than in the
field, explaining why variations in the mixture of types do not lead to
significant differences in the cluster LFs. The differences between the
field and cluster LFs for the various spectral types can be
qualitatively explained by the suppression of star formation in the
dense cluster environment, together with mergers to produce the
brightest early-type galaxies.
\\
\\
\noindent
{\bf Key words:}\quad galaxies: luminosity function, mass function -- galaxies: formation
\end{abstract}

\newpage
\strut\newpage
\strut\newpage

\section{Introduction}

The galaxy luminosity function (hereafter LF), which describes the
number of galaxies per unit volume as a function of luminosity, is a
fundamental tool for testing theories of galaxy formation and
interpreting observations of galaxies at high redshift for evidence of
evolution. Furthermore, precise and accurate measurements of the LF in
different environments have the potential to provide important clues as
to the role of `environmental' processes (e.g., dynamical interactions
in rich clusters) in determining the properties of the present-day
galaxy population.

One of the main legacies of the numerous redshift surveys that have been
undertaken in the last decade or more is the wealth of LF measurements
for galaxies in the low density `field' environment. A compilation and
comparison of these various measurements was recently published by
\citet{cro01}. This work showed that while the data were adequately
represented by a Schechter function, there were serious discrepancies
between the various measurements, with the LFs differing by as much as a
factor of 2 at the $L^{*}$ point, and with a scatter of a factor of 10
at $0.01L^{*}$. These differences are due to a combination of surface
brightness selection, colour, aperture effects and local density
variations among others. These problems have now been overcome in the
recent analysis of large redshift surveys such as the 2dF Galaxy
Redshift Survey (2dFGRS: Folkes et al. 1999, Madgwick et al. 2001,
Norberg et al. 2002) and the Sloan Digital Sky Survey (SDSS: Blanton et
al. 2001).

Rich clusters of galaxies provide the other extreme in environment,
representing the highest density regions inhabited by galaxies. It is
generally easier to derive LFs in clusters, since they provide rich
ensembles of galaxies all at the same distance, whose over-density with
respect to the surrounding field is sufficiently high to efficiently
identify members either photometrically, through the statistical removal
of foreground and background galaxies (e.g., Driver et al. 1998a and
references therein) or spectroscopically \citep{sma97,dep98,ada00}

These techniques have been used to measure LFs for individual clusters,
which in many cases have been combined to form a `composite' LF to
improve statistics (particularly at the brightest luminosities) and
average out systematic uncertainties \citep{dre78, lug86, col89, lug89,
gai97, lum97, val97, ram98, gma99, pao01, got02, yag02}. However, these
studies have not been unanimous on the exact form of the LF, with some
claiming there to be significant differences between the LFs from
cluster to cluster and between cluster and field \citep{dre78, lpc97,
lum97, val97, gma99, dri98a, got02}, while others finding no differences
and concluding that galaxies in all environments appear to be drawn from
a single, `universal' LF \citep{lug86, col89, lug89, gai97, ram98,
tre98, pao01, yag02}. A summary of the Schechter function fits from some
of these previous studies (all of which are based on the technique of
background subtraction, unlike the present work which uses spectroscopic
identifications of cluster members) is given in Table 1. We have
transformed magnitudes to $H_0=100$ km s$^{-1}$ Mpc$^{-1}$ but we have
not changed their cosmology.

\begin{table*}
\begin{center}
\begin{minipage}{140mm}
\caption{Composite LFs for rich clusters in blue passbands.}
\begin{tabular}{ccccc}
\hline
Reference & M$^*$ & $\alpha$ & N$_{\rm clusters}$ & Luminosity Range \\ 
\hline
Schechter (1976) & $-19.9 \pm 0.5$ & $-1.24$ & 13 & $-22.5 < M_J < -18.5$ \\
Dressler (1978) & $-19.7 \pm 0.5$ & $-1.25$ & 12 &$ -23.5 < M_F < -18.5$ \\
Colless (1989) & $-20.10 \pm 0.07$ & $-1.25$ & 14 & $-22.5 < M_J < -17$ \\
Lumsden et al. (1997)\footnote{Average of three methods} & $-20.16 \pm 0.02$ & $-1.22 \pm 0.04$ & 46 & $-21 < M_b < -18$ \\
Valotto et al. (1997) & $-20.0 \pm 0.1$ & $-1.4 \pm 0.1$ & 55 & $-21 < M_b < -17$ \\
Rauzy et al. (1998) & $-19.91 \pm 0.21$ & $-1.50 \pm 0.11$ & 28 & $-21 < M_b < -17$ \\
Garilli et al. (1999) & $-20.30 \pm 0.10$ & $-0.94 \pm 0.07$ & 65 & $-22.5 < M_g < -15.0$ \\
Paolillo et al. (2001) & $-20.22 \pm 0.15$ & $-1.07 \pm 0.08$ & 39 & $-24.5 < M_g < -16.5$ \\
Goto et al. (2002) & $-21.24 \pm 0.11$ & $-1.00 \pm 0.06$ & 204 & $-24 < M_{g'} < -17$ \\
\hline
\end{tabular}
\end{minipage}
\end{center}
\end{table*}

With the 2dF Galaxy Redshift Survey (2dFGRS; Colless et al. 2001) now
complete, there is an opportunity to revisit these issues and
simultaneously address the detailed form of the LF in rich clusters and
in the field in a consistent manner. High-quality measurements of the
field LF based on 2dFGRS data have already been published by
\cite{mad02} and \cite{nor02}. In this paper we present an analysis of
the LFs of 60 rich clusters, taken from the sample of known clusters
within the 2dFGRS survey region that were identified and characterised
by \cite{dep02}.

The enormity of 2dFGRS in terms of its size, depth and sky coverage has
a number of distinct advantages in comparison to previous LF studies.
Firstly, cluster membership is determined unambiguously from
spectroscopic redshifts for nearly all galaxies, eliminating the
introduction of systematic errors into the derived LFs through field
subtraction \citep{dri98b}. Secondly, the apparent magnitude limit of
2dFGRS ($b_{J}=19.45$) is sufficiently deep that, at the redshifts
covered here ($z<0.11$), our study extends to $\sim 5$ magnitudes
fainter than $M^{*}$ -- at least as deep as previous studies. Thirdly,
the sheer number of clusters (60) that we can study together with the
almost 1-in-1 sampling of their galaxy populations, provides the level
of statistical discrimination needed, particularly at the bright end of
the LF \citep{col89}, to detect differences that are of physical
interest. Finally, our comparison of the cluster and field LFs is done
entirely within the 2dFGRS and hence based on the same input catalogue,
galaxy photometry and redshift observations. The field and cluster
samples therefore share most of the selection effects and observational
biases and the resulting LFs can be compared fairly: we also note that
the field LF in \cite{mad02} is derived from galaxies with $z < 0.15$
and is therefore similar to the volume-limited sample of cluster
galaxies.

The plan of this paper is as follows: The next section describes our
cluster sample and the procedure used for constructing composite LFs. We
then present our derived LFs in Section 3, both for the entire sample of
clusters and for subsets differentiated on the basis of velocity
dispersion, Bautz-Morgan type, richness, and the presence of
substructure. Section 4 compares our data with previous work and the
field. Our results are discussed and summarised in Section 5. We adopt
the `concordance' cosmology with $\Omega_m=0.3$, $\Omega_{\Lambda} =0.7$
and $H_0=100$. This is about 0.07 magnitudes brighter than the
Einstein-DeSitter model at the mean redshift ($z=0.07$) of our cluster
sample.

\section{Cluster selection and LF construction} 

\subsection{Cluster sample}

The clusters studied here were drawn from the sample of known clusters
within the 2dFGRS, constructed by cross-matching the Abell
\citep{abe58,aco89}, APM \citep{dal97} and EDCC \citep{lum92} catalogues
with the 2dFGRS catalogue \citep{dep02}. Our selection was restricted to
clusters with $z < 0.11$ -- in order to sample well below the predicted
$M^*$ -- and those with at least 40 confirmed members within the Abell
radius (1.5\,$h^{-1}$\,Mpc). As part of our previous investigation of
the redshifts and velocity dispersions of these clusters \citep{dep02},
we used a `gapping' algorithm to identify their {\it bona fide} members.
It is these galaxies that we use for our LF construction. Table 2 lists
the clusters used in this study and their relevant properties. We only
present a few lines here, the full table being available electronically.
The total number of clusters used is 60. As per \cite{dep02}, only
unique names are cited in the ID column in order of preference as Abell,
APM and EDCC; cross references to other cluster catalogues are available
in a table on the WWW at the Astrophysical Data Center
(http://adc.gsfc.nasa.gov/adc.html). We stress that this is not intended
as an homogeneous sample, selected according to well-defined criteria
(e.g., X-ray luminosity,) but a selection of the richest nearby
clusters, although the sample is likely to be complete in this respect.
The 4186 galaxies used here make up about 3\% of the total 2dF sample at
$ z < 0.11$.

For clusters not classified by \cite{aco89}, we determine a Bautz-Morgan
(B-M) type based on the luminosity distribution of the brightest members.
Our cluster database upon which the \cite{dep02} study was based, has
since been updated to reflect the final survey total of 221,000
galaxies.

\begin{table*}
\begin{center}
\caption{Clusters studied in this paper.}
\begin{tabular}{cccccccc}
\hline
Cluster ID & RA(1950) & Dec.(1950) & B-M Type & $cz$ & $\sigma$ & N$_{\rm members}$ & Completeness \\ 
\hline
A0930   & 10:04:30.65 & $-$05:22:48.4 &    III & 17316 &  907 &  91 & 0.84 \\
A0954   & 10:11:11.10 & $+$00:07:40.2 &     II & 28622 &  832 &  49 & 0.72 \\
A0957   & 10:11:05.10 & $-$00:40:38.9 &   I-II & 13623 &  722 &  88 & 0.71 \\
A1139   & 10:55:36.68 & $+$01:52:20.9 &    III & 11876 &  504 & 106 & 0.82 \\
A1189   & 11:08:30.14 & $+$01:21:42.6 &    III & 28824 &  814 &  42 & 0.77 \\
A1200   & 11:10:03.25 & $-$02:56:27.6 &    III & 24970 &  825 &  62 & 0.83 \\
A1236   & 11:20:10.82 & $+$00:44:10.0 &     II & 30533 &  589 &  41 & 0.83 \\
A1238   & 11:20:20.36 & $+$01:23:19.4 &    III & 22160 &  586 &  85 & 0.82 \\
A1248   & 11:21:08.28 & $-$03:56:31.4 &   I-II & 16139 &  798 &  44 & 0.82 \\
A1364   & 11:40:55.99 & $-$01:27:52.8 &    III & 31859 &  600 &  51 & 0.85 \\
A1620   & 12:47:29.78 & $-$01:16:07.1 &    III & 25513 & 1095 &  95 & 0.89 \\
A1663   & 13:00:18.05 & $-$02:14:57.7 &     II & 24827 &  884 &  91 & 0.80 \\
A1692   & 13:09:41.25 & $-$00:39:59.7 & II-III & 25235 &  686 &  65 & 0.80 \\
A1750   & 13:28:36.52 & $-$01:28:15.9 & II-III & 25647 &  981 &  78 & 0.62 \\
A2660   & 23:42:39.94 & $-$26:06:42.5 &   I-II & 15974 &  845 &  52 & 0.67 \\
A2716   & 00:00:27.51 & $-$27:24:50.3 &   I-II & 19889 &  660 &  60 & 0.84 \\
A2734   & 00:08:49.47 & $-$29:07:58.1 &    III & 18646 & 1038 & 127 & 0.91 \\
A2780   & 00:27:35.61 & $-$29:44:02.1 &    III & 29987 &  782 &  46 & 0.83 \\
A3027   & 02:28:42.26 & $-$33:19:27.7 &    III & 23166 &  907 &  91 & 0.65 \\
A3094   & 03:09:16.42 & $-$27:07:08.4 &    III & 20475 &  774 & 107 & 0.84 \\
A3880   & 22:25:04.97 & $-$30:49:51.5 &     II & 17258 &  840 & 122 & 0.83 \\
A4012   & 23:29:11.30 & $-$34:19:50.8 & II-III & 16230 &  498 &  73 & 0.88 \\
A4013   & 23:27:42.53 & $-$35:13:21.6 &    III & 16410 &  904 &  85 & 0.81 \\
A4038   & 23:44:59.00 & $-$28:24:10.0 &    III &  9077 &  933 & 175 & 0.89 \\
A4053   & 23:52:10.66 & $-$27:57:34.6 &    III & 20927 &  994 &  58 & 0.93 \\
S0003   & 00:00:37.68 & $-$28:09:24.7 &      I & 19293 &  833 &  46 & 0.91 \\
S0006   & 00:02:09.11 & $-$30:45:42.7 &      I &  8768 &  630 &  53 & 0.90 \\
S0084   & 00:46:57.51 & $-$29:47:33.7 &      I & 32664 &  807 &  44 & 0.69 \\
S0141   & 01:11:26.28 & $-$32:00:45.6 &      I &  5793 &  411 & 110 & 0.74 \\
S0160   & 01:27:54.67 & $-$33:09:41.4 &      I & 20638 &  627 &  54 & 0.90 \\
S0166   & 01:32:06.97 & $-$31:51:43.3 &     II & 20908 &  511 &  50 & 0.99 \\
S0167   & 01:32:06.91 & $-$33:05:30.3 &      I & 19792 &  769 &  66 & 0.89 \\
S0258   & 02:23:33.21 & $-$29:50:26.9 &     II & 18026 &  593 &  87 & 0.72 \\
S0301   & 02:47:27.22 & $-$31:23:46.7 &      I &  6652 &  608 &  92 & 0.82 \\
S0333   & 03:13:04.34 & $-$29:25:41.3 &     II & 20042 &  998 &  74 & 0.90 \\
S0340   & 03:17:55.68 & $-$27:11:45.6 & II-III & 20281 &  939 &  43 & 0.87 \\
S1043   & 22:33:43.18 & $-$24:36:05.2 &      I & 11091 & 1345 & 116 & 0.79 \\
S1086   & 23:02:06.51 & $-$32:49:14.8 &   I-II & 25561 &  509 &  53 & 0.87 \\
S1136   & 23:33:38.55 & $-$31:52:48.8 &    III & 18688 &  617 &  50 & 0.82 \\
S1142   & 23:38:17.18 & $-$30:32:47.5 & II-III & 24425 &  669 &  40 & 0.89 \\
S1165   & 23:55:24.91 & $-$30:08:40.9 &   I-II &  8920 &  359 &  56 & 0.90 \\ 
S1171   & 23:58:45.20 & $-$27:41:54.5 &     II &  8763 &  788 &  53 & 0.90 \\
APM039  & 00:14:26.04 & $-$31:38:15.7 &    III & 31792 &  559 &  58 & 0.94 \\
APM078  & 00:27:44.53 & $-$29:53:26.8 &    III & 30079 &  796 &  41 & 0.83 \\
APM268  & 02:27:48.57 & $-$33:23:55.5 &    III & 23223 &  832 &  97 & 0.62 \\
APM917  & 23:38:58.49 & $-$29:30:49.6 &      I & 15358 &  503 &  77 & 0.89 \\
APM945  & 23:56:27.74 & $-$32:04:33.3 &      I & 17982 &  536 &  41 & 0.87 \\
APM954  & 23:58:20.95 & $-$28:44:30.4 &    III & 18475 &  445 &  67 & 0.91 \\
EDCC069 & 21:55:50.75 & $-$28:42:15.8 &   I-II &  6528 &  528 &  64 & 0.72 \\ 
EDCC119 & 22:13:32.57 & $-$25:55:10.7 &      I & 25546 & 1112 &  43 & 0.84 \\
EDCC142 & 22:22:50.88 & $-$31:27:17.9 &   I-II &  8411 &  274 &  44 & 0.77 \\
EDCC142 & 22:22:45.48 & $-$31:18:50.8 & II-III & 17437 &  296 &  53 & 0.75 \\
EDCC153 & 22:29:25.49 & $-$31:29:12.8 &    III & 17452 &  708 &  40 & 0.69 \\
EDCC155 & 22:29:22.08 & $-$25:39:20.6 &   I-II & 10310 &  714 &  52 & 0.82 \\
EDCC365 & 23:52:33.51 & $-$33:01:07.7 &     II & 17748 &  524 &  79 & 0.83 \\
EDCC442 & 00:23:02.22 & $-$33:19:24.5 &    III & 14867 &  763 & 127 & 0.88 \\
EDCC457 & 00:33:35.03 & $-$26:22:00.2 &    III & 18492 &  977 &  66 & 0.86 \\
EDCC652 & 02:25:11.88 & $-$29:51:00.7 & II-III & 17944 &  583 &  64 & 0.81 \\
EDCC661 & 02:29:45.79 & $-$32:11:30.6 &    III & 24362 &  302 &  48 & 0.86 \\
EDCC664 & 02:30:56.67 & $-$33:02:58.9 &   I-II & 23750 &  399 &  52 & 0.75 \\
\hline
\end{tabular}
\end{center}
\end{table*}

\subsection{Composite LFs}

The quality of individual cluster LFs varies, depending on the number of
members and the completeness of the redshift identification as described
in greater detail below. Rather than present LFs for each individual
cluster, we derive a `composite' LF and study its variation in
subsamples constructed according to physically meaningful criteria
(e.g.\ cluster mass, dynamical evolutionary status). This approach makes
it feasible to look for real differences which are hidden by small
number statistics in individual cases.

Composite LFs were built following the prescriptions of \cite{col89}, by
summing galaxies in absolute magnitude bins and scaling by the richness
of their parent cluster. Specifically, the following summation was
carried out:
\begin{equation}
N_{cj}={N_{c0} \over m_j}{\sum_i} {N_{ij}  \over N_{i0}} ~,
\end{equation}
where $N_{cj}$ is the number of galaxies in the $j$th absolute magnitude
bin of the composite LF, $N_{ij}$ is the number in the $j$th bin of the
$i$th cluster LF, $N_{i0}$ is the normalization used for the $i$th
cluster LF (taken as the corrected number of galaxies brighter than
$M_{b_{\rm J}}=-19$; see below for details), $m_j$ is the number of
clusters contributing to the $j$th bin, and $N_{c0}$ is the sum of all
the normalizations:
\begin{equation}
N_{c0}={\sum_i} N_{i0} ~.
\end{equation}

The formal errors in $N_{cj}$ are computed according to:
\begin{equation}
\delta N_{cj} = {N_{c0} \over m_j} {\Big [ \sum_i \Big (} {\delta
N_{ij} \over N_{i0}} {\Big)} ^2 {\Big ]}^{1/2} ~,
\end{equation}
where $\delta N_{cj}$ and $\delta N_{ij}$ are the formal errors in the
$j$th LF bin for the composite and $i$th cluster, respectively.

For each cluster we count galaxies in bins of absolute magnitude:
\begin{equation}
M_{b_{\rm J}}=b_{\rm J}-\mu-A_{b_{\rm J}}-K_{b_{\rm J}}^{z} ~,
\end{equation}
where $b_{\rm J}$ is the apparent magnitude, $\mu$ is the distance
modulus, $A_{b_{\rm J}}$ is the extinction and $K_{b_{\rm J}}^{z}$ is
the $k-$correction. In principle, the $k-$correction could be determined
from the $\eta$ parameter \citep{mad02}, but this is only possible for a
fraction of the cluster members with adequate signal-to-noise ratio in
their spectra. For this reason we adopt a single $k-$correction of the
form
\begin{equation}
K_{b_{\rm J}}^{z} = 2.6z + 4.3z^2 ~,
\end{equation}
following \cite{mad02}, which is appropriate for the early-type galaxies
which predominate amongst our sample of cluster members (Fig.~1). The
average offset between this correction and the one for star forming
galaxies is 0.09 magnitudes.

\begin{figure}
\includegraphics[width=84mm]{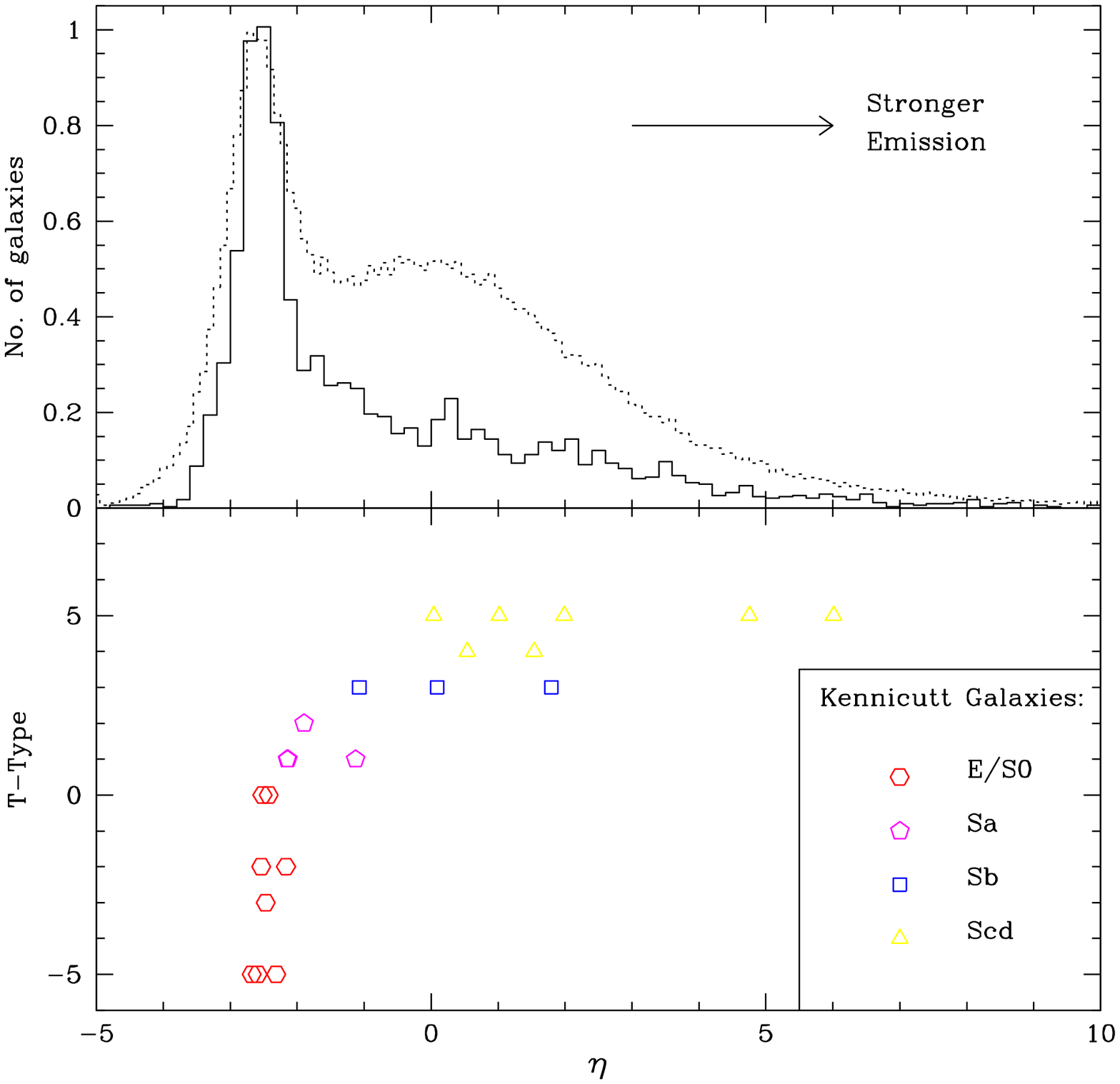}
\caption{ 
Upper panel: Distribution of $\eta$ parameter values for galaxies in our
clusters (solid line) and in the general field (dotted line). Lower
panel: The relationship between morphology (represented by T-type) and
$\eta$ value as defined by Kennicutt (1992) for nearby galaxies;
different symbols are used to indicate the corresponding Hubble types
(see legend). It can be seen that early-type galaxies ($\eta \sim -2.5$)
dominate the cluster populations. The relative proportions of type
1,2,3,4 galaxies are 0.54,0.24,0.13,0.09 in clusters as opposed to
0.36,0.32,0.20,0.12 in the field, where types are defined as in Madgwick
et~al.\ (2002) and in the text below. }
\end{figure}

\begin{figure}
\centering\includegraphics[width=86mm]{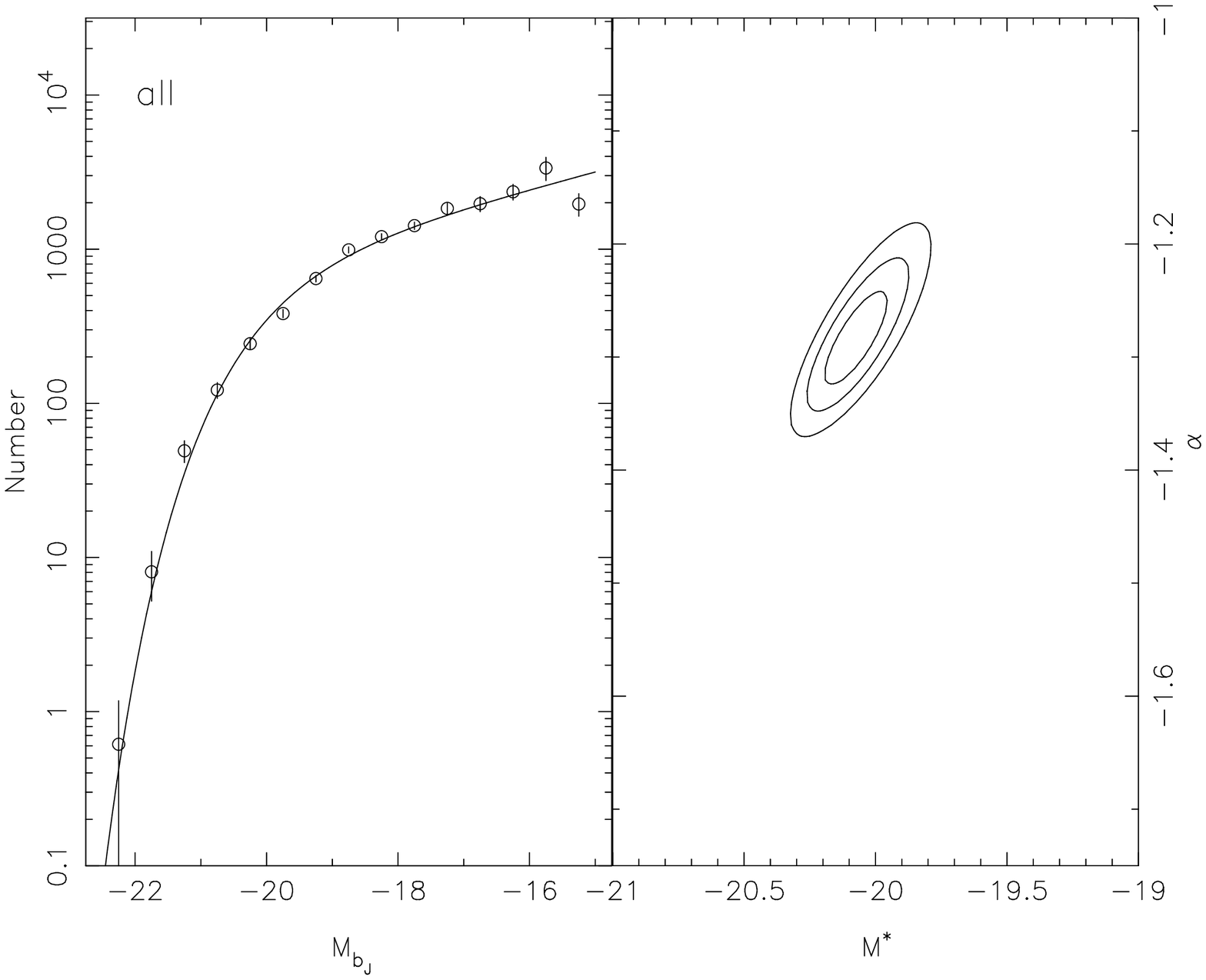}
\caption{ 
Left panel: The luminosity function of cluster galaxies for the whole
sample ({\it open circles}), showing the best Schechter function fit
(solid line) with parameters as in the text and Table 3. Right panel: 1,
2 and 3$\sigma$ error contours for the LF. }
\end{figure}

Since we do not have redshifts for all galaxies in the cluster fields,
(the overall completeness is 81\%) we correct for this as a function of
apparent magnitude. In each apparent magnitude bin (corresponding to the
appropriate absolute magnitude bin for each cluster) we count the number
of cluster members $N_C$, the number of galaxies with a measured
redshift $N_R$ and the number of galaxies in the input catalogue $N_I$.
The completeness-corrected number of cluster members in each bin is
therefore
\begin{equation}
N_{ij}={{N_C N_I}\over{N_R}} ~,
\end{equation}
which assumes that redshift identification is not biased towards or
against galaxies that are members of rich clusters. The size of this
correction varies from cluster to cluster, and as a function of
magnitude, but is typically small ($\sim$10--20\%).

Here $N_I$ is a Poisson variable, since it is drawn from an ideal
(infinite) distribution, and $N_C$ is a binomial random variable, the
number of `successes' (cluster members) in $N_R$ `trials' (redshift
measurements) with probability of success $N_C/N_R$. Therefore the
errors are given by
\begin{equation}
{{\delta^2 N_{ij}} \over {N^2_{ij}}}={{\sigma^2 (N_I)} \over N^2_I}+
{{\sigma^2 (N_C)} \over N^2_C} ~,
\end{equation}
which simplifies, using the standard binomial error expression, to
\begin{equation}
{{\delta^2 N_{ij}} \over {N^2_{ij}}}
           = {1\over N_I}+{1\over N_C}-{1\over N_R} ~.
\end{equation}

We also determine LFs for different galaxy spectral types. These types
are the same as those used by \cite{mad02} in their analysis of the
field LF and are based on a classification parameter ($\eta$) derived
from a principal component analysis of the 2dFGRS spectra. We determine
LFs for the early (type 1), mid (type 2) and late (types 3+4) spectral
classes (there are too few galaxies to compute separate LFs for the
latter two classes). Fig.~1 shows the relationship between spectral type
and morphological type and also the distribution of spectral type in
both the cluster and field samples. The types are defined as in
\cite{mad02}: type 1 galaxies have $\eta < -1.4$, type 2's $-1.4 < \eta
< 1.1$ and types 3 and 4 $\eta > 1.1$.

The composite luminosity functions for each spectral type are computed
as for the overall sample, except that the completeness-corrected number
of galaxies is determined separately for each type as 
\begin{equation}
N_{ij}(S)= N_{ij} \times {N_C (S) \over {\sum_{S} N_C (S)}} ~,
\end{equation}
where $N_C (S)$ is the number of cluster galaxies having spectral type S
(1, 2 or 3+4) and the summation is carried out for cluster members over
all spectral types. The errors are given by
\begin{equation}
{\delta^2 N_{ij} (S) \over {N^2_{ij} (S)}}={1\over N_I}+
{1\over {N_C (S)}}-{1\over N_R},
\end{equation}
where $N_C (S)$ is a binomial variable and ${\sum_{S} N_C (S)}$ is a
Poisson variable, with $k-$corrections for each spectral type as given
in \cite{mad02}.

\section{Results}

In Fig.~2, we show the composite LF derived for our complete ensemble of
60 clusters over the full range of absolute magnitude: $-22.5 <
M_{b_{\rm J}} < -15$. This LF is based on 4,186 cluster members,
yielding errors less than 2\% over the range $-21.0 < M_{b_{\rm J}} <
-16.0$. A $\chi^2$ fit of this LF by a \cite{sch76} function shown in
Fig.~2 as a solid curve gives a characteristic magnitude of $M^*_{b_{\rm
J}}=-20.07 \pm 0.07$ and a faint-end power-law slope of $\alpha=-1.28
\pm 0.03$. We tabulate these values in Table 3 together with a measure
of the goodness of fit. This is only good to about 1\% because of the
large discrepancies with the last two bins, where completeness
corrections are larger and only a few (4) clusters contribute.

Here, as for all our other LFs, the $1\sigma$ errors are derived by a
Monte Carlo simulation, where the independent variable vector is
replicated 1,000 times, each data point being replaced by the fitted
function value plus noise (based on the original error bar for each
point), and the function fitted again to the simulated data. The
dispersion about the derived $M^*$ and $\alpha$ is then used to
calculate the errors. The right-hand panel of Fig.~2 shows the $\chi^2$
error contours corresponding to the 1, 2 and 3$\sigma$ levels. It can be
seen the $M^*$ and $\alpha$ values are highly correlated, and hence the
errors in these quantities quoted in Table 3 are not independent. The
fitting procedure is carried out over all three parameters and we
marginalized over the normalization to draw these error contours.

We then consider LFs split according to the spectral types described in
\cite{mad02}. These types are closely related to the star formation rate
and somewhat more loosely to morphology. Our purpose here is to consider
the universality of type-specific LFs and how the cluster environment
affects star formation. Fig.~3 shows the corresponding data and fits,
with error ellipses, for galaxies of spectral types 1, 2 and 3+4.

\begin{figure*}
\includegraphics[width=120mm]{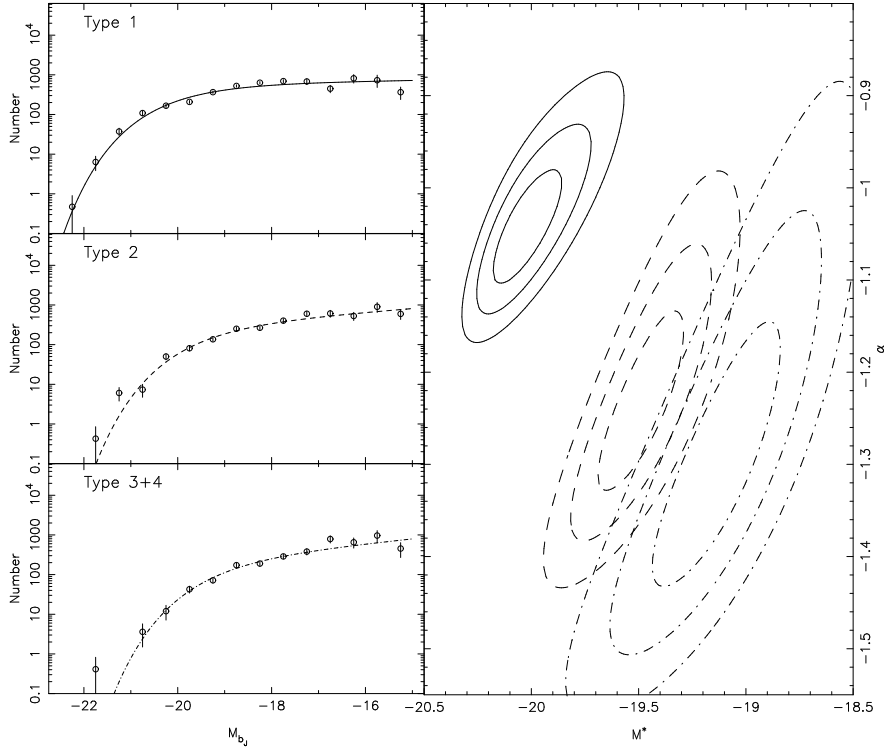}
\caption{
Left panels: Luminosity functions for type 1, 2 and 3+4 galaxies (as
identified in the panel legends), together with their best-fitting
Schechter functions. Right panel: 1, 2 and 3$\sigma$ error contours for
the Schechter function fits (with the same line styles as in the
lefthand panels). }
\end{figure*}

For purposes of comparison, and to assess the effect of removing the
brightest cluster galaxies, we also carry out a fit to the data in
Fig.~2 over a restricted absolute magnitude range, $M_{b_{\rm J}} >
-21.5$. This yields a LF with $M^*_{b_{\rm J}}=-20.02 \pm 0.14$ and
$\alpha = -1.27 \pm 0.04$---exclusion of the brightest cluster members
thus has no effect on the faint-end slope and leads to a marginally
fainter characteristic magnitude, as expected.

We also consider cluster subsamples and plot their LFs and error
contours as described below:

(i) We split the sample at a velocity dispersion of $800$\,km\,s$^{-1}$,
as this is the approximate value at which the distribution of cluster
velocity dispersions, N({$\sigma$}), turns over \citep{dep02},
separating massive systems from relatively poor ones. About the same
number of galaxies are found in each subsample.

(ii) We used the normalisation parameter, $N_{i0}$, as a measure of
richness, with its median value being used to divide the sample. Here
the choice is somewhat arbitrary and is mostly motivated by having
similar numbers of galaxies in each subset and hence equal statistical
weights.

(iii) Bautz-Morgan (B-M) types measure the dominance of the brightest
cluster members over the rest of the cluster population. In the
cannibalism model of \cite{osha78}, the brightest cluster galaxies grow
at the expense of other less massive galaxies and therefore the LF
should vary as the cluster evolves. We chose to divide clusters into two
bins: `early' B-M types, (B-M types I, I-II and II) and `late' B-M types
(having B-M types II-III and III). The former may be evolved systems,
whereas the latter may be at an earlier stage of dynamical evolution.
Again, this split ensures approximately equal numbers of galaxies in
each composite LF.

(iv) The presence of substructure may be an indicator of recent or
on-going cluster merging: substructure can be measured from the 3D
distribution of cluster members, although to some extent the definition
of substructure is arbitrary. We use the Lee statistic first applied to
clusters by \cite{fit88}, which measures the probability that cluster
galaxies form a single group or can be split into two groups in
position-velocity space. Since most clusters showing substructure
consist of two groups this approach is economical. We arbitrarily
defined clusters with substructure to be those where the Lee statistic
indicates a greater than 50\% probability that they consist of two
groups. By this definition, about 25\% of our clusters contain
substructure, a fraction comparable to the 31\% of ENACS clusters
showing substructure \citep{ssg99}.

(v) We analyze radial trends by considering galaxies inside and outside
two King core radii from the cluster centre. The King core radius is
about 150\,$h^{-1}$\,kpc \citep{ada98} and, as the densest region of the
cluster, is the most likely to show environmental effects. Our choice of
two King radii is again motivated by the need to obtain sufficient
statistics (i.e.\ about half the galaxies in each sample).

LFs for all these subsamples, together with their best-fitting Schechter
functions and the associated error ellipses are shown in Fig.~4.

\begin{figure*}
\includegraphics[width=180mm]{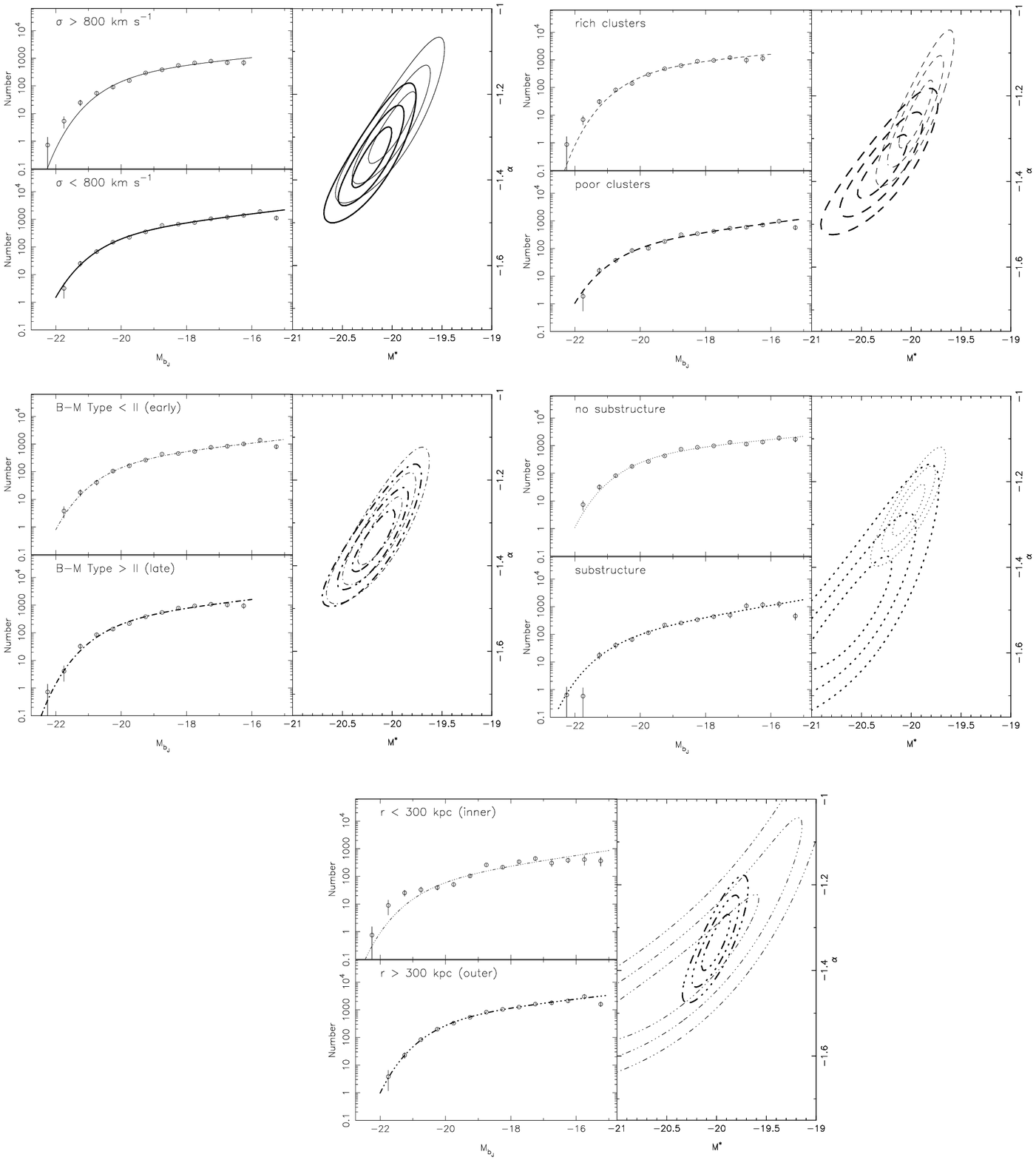}
\caption{
Luminosity functions for each cluster subsample, with best-fit Schechter
functions and error contours. The LFs are identified in the legends for
each panel and the error ellipses have the same line style as their
parent LFs. For example, the thin solid line in the upper left hand
panel shows the LF for galaxies in clusters with high velocity
dispersion, while the thick solid line is for clusters with low velocity
dispersion. }
\end{figure*}

Even with our large sample of clusters, some of the brighter and/or
fainter absolute magnitude bins are not well-populated in a few of the
above subsamples---the brighter bins because very bright galaxies are
intrinsically less common, and the fainter bins because of the
relatively small numbers of clusters at low redshift. For this reason we
carry out our fit over a smaller luminosity range, $-21.5 < M_{b_{\rm
J}} < -16.0$, for this part of the analysis. For purposes of comparison,
the total LF over the same range has $M^*_{b_{\rm J}}=-20.21 \pm 0.11$
and $\alpha=-1.36 \pm 0.04$.

Table~3 shows the derived values for $M^*$ and $\alpha$, and the
1$\sigma$ errors, derived from the Monte Carlo simulations referred to
above, for our full sample, the cluster subsamples described in the
previous section and the individual spectral types. We also tabulate
values for the field LF from \cite{mad02}. Data points for all our LFs
are also made available in electronic form on the MNRAS website.

\begin{table*}
\begin{center}
\caption{Summary of results.}
\begin{tabular}{ccccc}
\hline
Sample & $M^*_{b_{\rm J}}$ & $\alpha$ & $P(\chi^2>\chi^2_{\rm obs})$ & D/G \\ 
\hline
all ($-22.5<M_{b_{\rm J}}<-15.5$) & $-20.07 \pm 0.07$ & $-1.28 \pm 0.03$ & 0.013 & $2.09 \pm 0.18$ \\
all ($-21.5<M_{b_{\rm J}}<-16.0$) & $-20.21 \pm 0.11$ & $-1.36 \pm 0.04$ & 0.332 & $2.09 \pm 0.18$ \\
Type 1                            & $-20.04 \pm 0.09$ & $-1.05 \pm 0.04$ & 0.000 & $1.29 \pm 0.16$ \\
Type 1 (rich)                     & $-20.14 \pm 0.12$ & $-1.22 \pm 0.06$ & 0.156 & $1.41 \pm 0.38$ \\
Type 1 (poor)                     & $-20.06 \pm 0.17$ & $-1.02 \pm 0.07$ & 0.001 & $1.28 \pm 0.14$ \\
Type 2                            & $-19.48 \pm 0.13$ & $-1.23 \pm 0.07$ & 0.606 & $2.68 \pm 0.53$ \\
Type 3+4                          & $-19.14 \pm 0.19$ & $-1.30 \pm 0.10$ & 0.182 & $4.88 \pm 0.97$ \\
$\sigma > 800$                    & $-19.99 \pm 0.16$ & $-1.28 \pm 0.05$ & 0.009 & $1.84 \pm 0.23$ \\
$\sigma < 800$                    & $-20.20 \pm 0.14$ & $-1.35 \pm 0.04$ & 1.000 & $2.11 \pm 0.25$ \\
rich                              & $-19.96 \pm 0.12$ & $-1.25 \pm 0.05$ & 0.083 & $1.68 \pm 0.36$ \\
poor                              & $-20.29 \pm 0.18$ & $-1.37 \pm 0.05$ & 0.233 & $2.10 \pm 0.15$ \\
B-M I,I-II,II                     & $-20.11 \pm 0.16$ & $-1.32 \pm 0.05$ & 0.000 & $2.12 \pm 0.21$ \\
B-M II-III,III                    & $-20.17 \pm 0.15$ & $-1.34 \pm 0.05$ & 0.002 & $1.83 \pm 0.28$ \\
no substructure                   & $-20.03 \pm 0.12$ & $-1.27 \pm 0.04$ & 0.003 & $1.85 \pm 0.17$ \\
substructure                      & $-20.59 \pm 0.31$ & $-1.51 \pm 0.08$ & 0.000 & $3.02 \pm 0.74$ \\
($r < 300$ kpc)                   & $-20.45 \pm 0.43$ & $-1.44 \pm 0.08$ & 0.000 & $2.02 \pm 0.36$ \\
($r > 300$ kpc)                   & $-19.83 \pm 0.08$ & $-1.29 \pm 0.05$ & 0.581 & $2.98 \pm 0.25$ \\ \hline
& Field \citep{mad02}& \\ \hline
all                               & $-19.79 \pm 0.04$ & $-1.19 \pm 0.01$ \\
Type 1                            & $-19.58 \pm 0.05$ & $-0.52 \pm 0.02$ \\
Type 2                            & $-19.58 \pm 0.03$ & $-0.96 \pm 0.01$ \\
Type 3                            & $-19.17 \pm 0.04$ & $-1.21 \pm 0.02$ \\
Type 4                            & $-19.19 \pm 0.04$ & $-1.36 \pm 0.03$ \\ 
Type 3+4                          & $-19.14 \pm 0.06$ & $-1.30 \pm 0.03$ \\
\hline
\end{tabular}
\end{center}
\end{table*}

\section{Discussion} 

We used the simulations of \cite{col89} to analyze the ability of our
data to detect differences in LF parameters: we find that our LFs are of
sufficient quality to detect differences of 0.15 mag in $M^*$ and 0.07
in $\alpha$ at the $1\sigma$ level, which is an improvement by a factor
of more than 2 over previous studies. Furthermore, unlike all previous
work, we are not limited by uncertainties in background subtraction. In
the following we compare our results to previous determinations of both
the cluster and field LFs and study the universality of the LF with
cluster subsamples. Type-specific LFs are also discussed. We finally
consider the implications of our findings for models of galaxy
formation.

\subsection{Comparison with Previous Work}

A comparison with previous work is shown in Table 1. Among previous
studies only \cite{gma99}, \cite{pao01} and \cite{got02} reach
luminosity limits comparable to ours. \cite{gma99} derive $M^*_g=-20.30
\pm 0.10$ and $\alpha=-0.94 \pm 0.07$. \cite{pao01} derive $M^*_g=-20.22
\pm 0.15$ and $\alpha=-1.07 \pm 0.08$ from their 39 clusters. Once we
correct for the colour difference between $g$ and $b_{\rm J}$ and for the
different cosmologies (using the mean redshifts of the cluster sample)
the characteristic luminosities of \cite{gma99} and \cite{pao01} agree
with our value within their 2$\sigma$ errors, whereas the two $\alpha$s
differ at about 2$\sigma$; the two LFs agree with each other at about
the 1.5$\sigma$ level: the difference is only marginally significant.

The LF of \cite{got02} from SDSS data is in considerable disagreement
with ours: $M^*$is brighter by about 1.2 magnitudes and $\alpha$ is
flatter. This LF disagrees with \cite{col89,lum97} and \cite{val97}, all
of which use APM data. It also disagrees with the $M^*$ and $\alpha$
values of \cite{gma99} and \cite{pao01} in the $g$ band. One possibility
is a systematic offset between APM and SDSS photometry, but this appears
to be ruled out by a comparison of SDSS Early Release Data, the
Millennium Galaxy Catalogue and 2dF photometry (Cross et al. 2002, in
preparation).

Earlier work is generally more limited in luminosity coverage: values
for the Schechter function fits for \cite{col89, lum92, val97} and
\cite{ram98} are, once we apply the necessary cosmological and filter
corrections, in reasonable agreement with our data. With the exception
of \cite{got02}, the agreement with previous work is generally
satisfactory.

\subsection{Cluster versus Field}

One of the motivations behind this work and previous studies is to test
the universality of the LF and its dependence on the environment. The
most dramatic comparison in this context is between rich clusters and
the field, because of the factor of 100 or more difference in galaxy
density between the two environments.

Having derived a statistically robust composite cluster LF from the
2dFGRS data, of foremost interest now is to make this comparison with
its 2dFGRS counterpart for the field. We base this comparison on the LFs
published by \cite{mad02}; these cover the absolute magnitude range
$-23.0 < M_{b_{\rm J}} < -13.5$, and include an overall field LF, and
LFs derived by dividing the galaxies into four different spectral
classes (see Figure 1). The $M^{*}_{b_{\rm J}}$ and $\alpha$ parameter
values for their Schechter function fits to these LFs are listed at the
bottom of Table 3. We also compare these LFs directly using a two-sample
$\chi^2$ test, the results of which, for both $M_{b_{\rm J}} < -16.0$
(the full magnitude range) and $M_{b_{\rm J}} < -18$ (the brighter
cluster members), are shown in Table 4.

\begin{table}
\begin{center}
\caption{$\chi^2$ comparisons of the field and cluster LFs.}
\begin{tabular}{ccc}
\hline
Sample & P($\chi^2$)         & P($\chi^2$)         \\ 
       & $M_{b_{\rm J}}<-18$ & $M_{b_{\rm J}}<-16$ \\
\hline  
All      & $<10^{-3}$ & $<10^{-3}$ \\
Type 1   & $<10^{-3}$ & $<10^{-3}$ \\
Type 2   & 0.133      & $<10^{-3}$ \\
Type 3+4 & 0.556      & 0.320      \\
\hline
\end{tabular}
\end{center}
\end{table}

The error ellipses for the Schechter function fits to all these LFs are
compared in Fig.~5. We only show the 3$\sigma$ contour for field
galaxies as the errors are small.

\begin{figure}
\includegraphics[width=84mm]{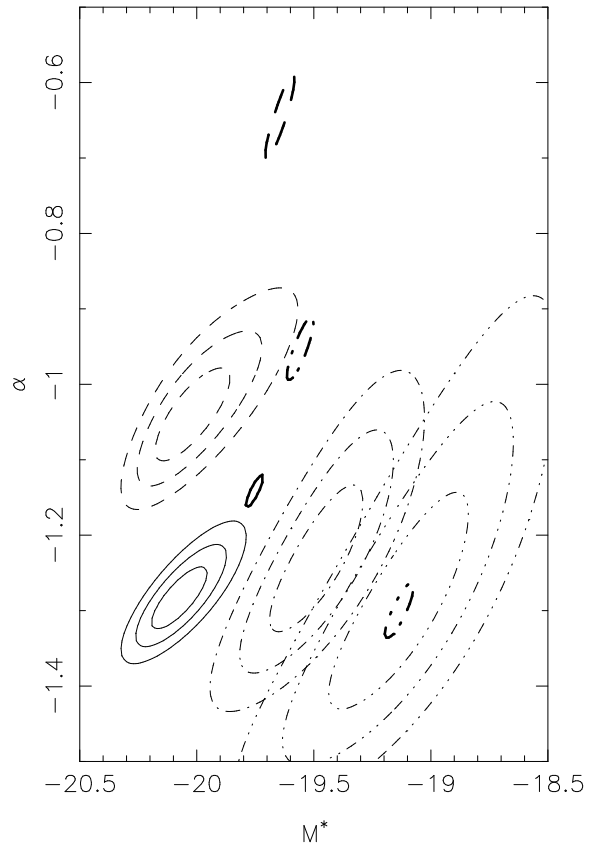}
\caption{ 
Error ellipses for the Schechter function fits to the field and cluster
LFs. The total LFs are represented by solid lines (thin for clusters and
thick for the field, as in Fig.~4). Dashed lines are for type 1
galaxies, dot-dashed lines for type 2 galaxies and dotted-dashed lines
for types 3+4. We only show the 3$\sigma$ error ellipse for the field
data. }
\end{figure}

The $\chi^2$ comparison shows that the overall cluster and field LFs
differ at more than 3$\sigma$, and this conclusion is also borne out by
a comparison of the error contours presented in Fig.~5. In terms of the
individual LF parameters, the overall cluster LF is 0.3 magnitudes
brighter in $M^*$ and 0.1 steeper in $\alpha$ than the overall field LF.
We note here that although the `field' LF is actually the total LF for
all environments, clusters do not contribute to the field sample to any
great extent, as cluster members are only 3\% of the total number of
galaxies in the survey; even for bright galaxies this fraction is less
than 6\%.

Significant differences are also found when the field and cluster LFs
for the different spectral classes are compared. For ease of comparison,
Fig.~6 shows type-dependent LFs for both field and cluster galaxies; the
the field LFs are normalized so that the overall field LF matches the
overall cluster LF at $M_{b_{\rm J}}=-18$.

\begin{figure*}
\includegraphics[width=160mm]{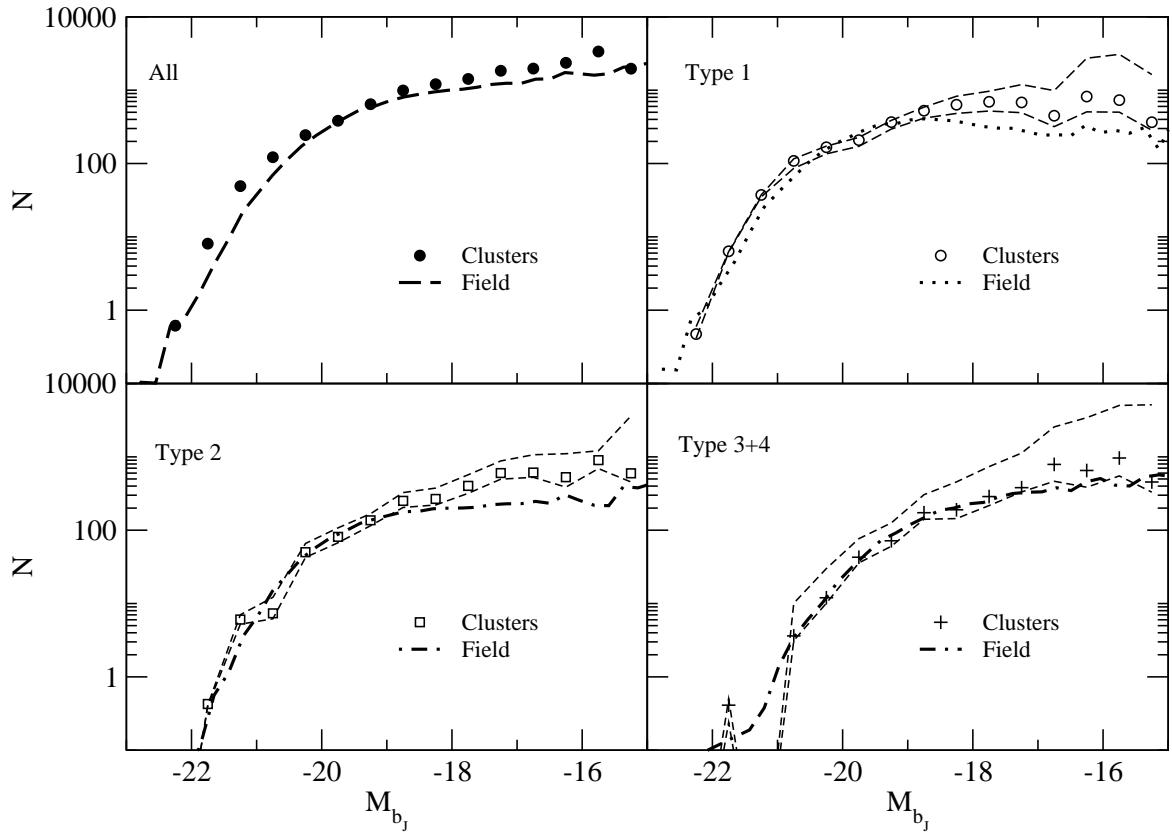}
\caption{ 
Comparison of cluster and field LFs. The open symbols are the
type-completeness corrected cluster counts shown in Figure 3. Thick
lines represent the field LFs from Madgwick et~al.\ (2002), normalized
to agree with the cluster LF at $M_{b_{\rm J}}=-19$. The two thin lines
show the luminosity distribution for no completeness correction (lower
line) and for the maximal correction discussed in the text (upper line).
Error bars are excluded for clarity (see Fig.~3 for error bars). The
error ellipses are shown in Fig.~5 for both the field and cluster
samples. }
\end{figure*}

The $\chi^2$ tests show that the LFs for type 1 cluster and field
galaxies are different over their entire range. This is due to the fact
that the type 1 LF in clusters is both brighter and much steeper than in
the field. The LFs of type 2 galaxies also differ, but Fig.~6 both shows
that the difference is primarily due to the steeper faint-end slope; the
$\chi^2$ fit shows that, to $M_{b_{\rm J}} < -18$, the two LFs match
adequately. Finally, the LF of star-forming galaxies (types 3 and 4) are
essentially identical in clusters and the field. These differences in
the shapes of the types-specific LFs, and the different contributions
from each type (as shown in Fig.~1), account for the differences between
the overall cluster and field LFs. The steeper faint-end slopes for type
1 and type 2 galaxies in clusters may lead, at magnitudes fainter than
those covered by our sample, to an upturn of the effective LF slope,
such as has been claimed to exist in some clusters \citep{dep98b}.

One caveat concerns the assumption implicit in equation~9, that the
galaxies without spectral types have the same type distribution as those
for which types could be determined. In order to test the importance of
this issue, we have computed LFs for each of the spectral types without
any completeness correction. We find that $M^*$ is hardly changed
whereas $\alpha$ is flattened by about 0.1; nevertheless, the LFs of
type 1 and 2 galaxies remain steeper than in the field whereas types 3+4
are consistent with the field LF. Incompleteness in spectral type comes
from two sources: galaxies which were not observed and therefore have
neither redshift nor spectral type and which are therefore an unbiased
sample, and galaxies whose spectra have too low a signal-to-noise to
yield a redshift or a spectral type. The untyped galaxies are roughly
divided evenly between these two categories. It is only the latter one
that may suffer from bias, in that galaxies of a specific type may be
preferentially misidentified. We have therefore adopted a `maximal
correction' where all of these galaxies are assigned to each type in
turn. This effect is small for types 1 and 2 but may be significant for
types 3 and 4. This is shown in Fig.~6. However, no matter what
correction is applied, cluster galaxies of all types have LFs with
faint-end slopes as steep or steeper than their counterparts in the
field.

These differences between the type-dependent LFs in the cluster and
field samples appear to be inconsistent with previous claims for a
universal type-dependent LF \citep{bst88,jeta97,and98}. However, it
needs to be noted that these earlier studies are concerned with
morphological types, which are only moderately well correlated with our
spectral types (Fig.~1), so that a given spectral type may include a
range of morphological types.

It is instructive at this stage to compare LF parameters for field and
cluster galaxies, divided by spectral type. Fig.~7 shows that,
progressing from late to early types, the cluster LFs have brighter
$M^*$ and flatter $\alpha$, as observed in the field LFs \citep{mad02}.
The trend in the value of $M^*$ with type is stronger in clusters than
in the field, with a similar value of $M^*$ for type 3+4 galaxies but a
significantly brighter $M^*$ for type 1 galaxies. However the trend of
$\alpha$ with type is weaker for cluster galaxies than for field
galaxies---at faint magnitudes the type 3+4 LF is steeply rising
($\alpha=-1.3$) in both clusters and the field, but the type 1 LF in the
field is actually falling at faint magnitudes ($\alpha=-0.52$) while in
clusters it is merely flat ($\alpha=-1.05$). This trend of $\alpha$ with
environment for type 1 galaxies is apparent even within the cluster
sample: the type 1 galaxies in richer clusters have a LF with a steeper
faint end than those in poorer clusters, which in turn have a steeper LF
than type 1 galaxes in the field (see Table~3 and Fig.~7). This is
consistent with the trend for star formation rate to be broadly
anticorrelated with the faint-end slope, which has been observed by
\cite{mar02} for galaxies in groups identified within the 2dFGRS 100K
release. Both \cite{yag02} and \cite{got02} claim similar variations in
their samples.

\begin{figure}
\includegraphics[width=84mm]{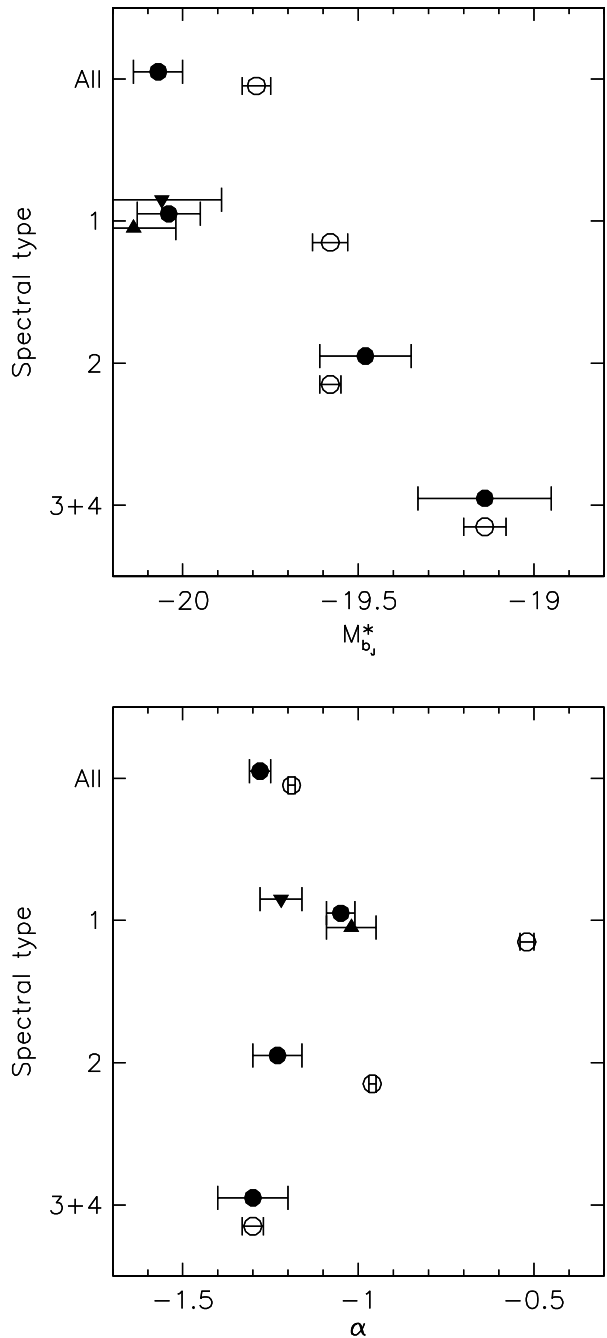}
\caption{
Comparison of cluster and field LF parameters. Open symbols are for
field galaxies and filled symbols for cluster galaxies; also shown are
the LF parameters for type 1 galaxies in richer clusters (triangle down)
and poorer clusters (triangle up). }
\end{figure}

\subsection{Cluster subsamples}

We have also compared subsamples of clusters chosen according to
velocity dispersion, richness, Bautz-Morgan type and likelihood of
containing substructure: similarly, we also considered samples of
galaxies within and outside two King core radii from the cluster centre.
These results are summarized in Table~3. We find that the values of
$M^*$ and $\alpha$ for each pair considered (e.g.\ low and high velocity
dispersion) do not differ at more than 1.5$\sigma$ in all cases. The
only possible exception is for the case of clusters containing
substructure, where $\alpha$ appears to be steeper. In all cases (again,
excepting clusters with likely substructure) the derived LF parameters
are consistent with those of the total LF within about 1.5$\sigma$. This
is in contrast with \cite{lum97}, who provided weak evidence for
differences in the LFs of clusters with high and low $\sigma$, and with
\cite{val97}, who suggested differences between rich and poor clusters.

The difference in the LF of clusters with substructure is potentially
interesting, as it is generally believed that substructure is an
indicator of recent or ongoing cluster merging. However an examination
of the confidence contours for the fitted Schechter function parameters,
shown in Fig.~4, shows that when the correlated nature of the parameters
is considered, the fits to the LFs of the clusters with and without
substructure are in fact consistent at better than 2$\sigma$.

The similarity of the LFs for the various subsamples is confirmed by
two-sample $\chi^2$ tests, which indicate that the probabilities that 
the pairs of contrasted subsamples have consistent LFs are: 99.3\% for 
early and late Bautz-Morgan clusters, 35.6\% for high and low $\sigma$ 
clusters, 50.0\% for rich and poor clusters, 27.1\% for clusters with 
and without substructure and 0.9\% for galaxies within and outside of 
300 kpc from the cluster centre. Inspection of Fig.~4 shows that the 
LFs of galaxies in the inner 300 kpc and the outer regions of clusters 
differ in detail; in particular, the inner region LF is a poor fit to 
the Schechter function, with a deficit of $L^*$ galaxies and an excess 
of brighter objects. This is reminiscent of the galactic cannibalism 
scenario of \cite{osha78} and \cite{muri84}, where $L^*$ galaxies are 
preferentially destroyed to fuel the growth of giant ellipticals, D 
and cD galaxies.

An alternate approach to quantifying the differences between LFs is to
calculate the dwarf to giant ratio (D/G). We define giants as galaxies
with $-21.5 < M_{b_{\rm J}} < -18$ and dwarfs as galaxies with $-18 <
M_{b_{\rm J}} < -16.0$, as in \cite{dri98a}. These ratios are tabulated
in Table~3 for all the LFs we consider. In general, our cluster samples
have D/G ratios comparable with those of moderately rich systems in
\cite{dri98a}. The D/G ratios for the contrasting pairs of subsamples
only differ at the $\sim$1$\sigma$ level, except for clusters with and
without substructure (1.5$\sigma$) and the inner and outer regions of
clusters (2.2$\sigma$)---in no case is the difference in D/G ratio
highly significant.

The overall impression is therefore one of broad universality of the LF
over a range of cluster properties. This is surprising if one considers
the very considerable differences between the LFs of the various
spectral types in the field \citep{mad02}, since different mixtures of
types would lead to different LFs. For instance, the morphology-density
relation might lead one to expect that rich and massive clusters would
be more elliptical-rich than poor, low-mass clusters; likewise, early
B-M clusters should be dominated by ellipticals, while late B-M type
clusters should be spiral-rich. Mixing the field LFs of the different
morphological mixes in these differing proportions would lead to
clusters with significantly different LFs. The reason this does not
appear to be the case is that the LFs of the different types are more
similar in clusters than they are in the field, so that changing the
mixture of types within clusters has less effect than one might have
expected.

\subsection{Implications for Galaxy Formation}

The main conclusions of our analysis are as follows:

(i) We have determined the composite LF of galaxies in clusters from the
2dFGRS. The LF is well-fitted by a Schechter function with parameters
$M^*_{b_{\rm J}}=-20.07 \pm 0.07$ and $\alpha=-1.28 \pm 0.03$. This is
significantly different to the field LF of \cite{mad02}, having a
characteristic magnitude that is approximately 0.3~mag brighter and a
faint-end power-law slope that is approximately 0.1 steeper.

(ii) There is no significant evidence for variations in the LF across a
broad range of cluster properties; the LF appears similar for clusters
with high and low velocity dispersions, for rich and poor clusters, for
clusters with early and late Bautz-Morgan types, and for clusters with
and without substructure. However the core regions of clusters differ
from the outer parts in having an excess of very bright galaxies.

(iii) Breaking down the LF by spectral type, the same trends are
apparent in clusters as in the field: the LFs of earlier-type galaxies,
with lower star-formation rates, have brighter characteristic magnitudes
and shallower faint-end slopes; the LFs of later-type galaxies, with
higher star-formation rates, have fainter characteristic magnitudes and
steeper faint-end slopes. The trend in faint-end slope, which is the
dominant difference between the LFs of the various spectral types in the
field, is much less pronounced in clusters. The smaller differences
between the LFs of different spectral types in clusters explain why
variations in cluster properties giving rise to significant variations
in the mixture of types do not lead to significant differences in the
cluster LFs.

(iv) A comparison between the field and cluster LFs of each spectral
type reveals that while the LF of late-type, star-forming galaxies is
very similar in clusters and the field, the LF of early-type galaxies
with low star-formation rates is both brighter and steeper in clusters
than in the field; intermediate types in clusters have an LF with a
similar bright end but a steeper faint end than their field
counterparts. As early and intermediate spectral types are predominant
in clusters, the overall cluster LF is also brighter and steeper than
the overall field LF.

The above results may be compared with two recent redshift-based studies
of galaxy clusters. \cite{dep98} derived the $K$-band LF of galaxies in
the inner 25~arcmin of the Coma cluster from a 100\%-complete sample of
members, and found that the cluster's LF is indistinguishable from the
general field LF. \cite{cz02} use a redshift survey of 6 clusters (with
a redshift completeness of 20--50\%) to derive $R$-band LFs for both the
cluster members and the field galaxies in the fore- and background of
the clusters. They also find that the $R$-band field and cluster LFs are
very similar. These results are for samples selected in red and infrared
passbands, and are therefore less sensitive to star formation, to which
our $b_{\rm J}$-selected sample is more closely correlated. In contrast
to these results, we find that there {\em are} small but statistically
significant differences between field and cluster galaxies, with the
blue-selected cluster LF being both brighter and steeper than that in
the field. The similarity between the field and cluster LFs in the red
and near-infrared passbands suggests that field and cluster galaxies
with the same total stellar mass have similar integrated star-formation
histories, while the difference between the field and cluster LFs in the
blue simply means that the star-formation rate is significantly affected
by the cluster environment.

Up until now it has been possible to claim that the observed differences
between cluster and field LFs were due in large part simply to the
different proportions in which supposedly universal type-specific LFs
were mixed in the two environments. This explanation is now precluded by
the finding that the type-specific LFs differ significantly with
environment (in fact, as Fig.~5 shows, the difference between clusters
and the field is less for the overall LF than for some of the
type-specific LFs).

Another way to understand the differences between the field and cluster
LFs is to consider a simple `closed box' model. In this model one
considers the type-specific field LFs to be the initial LFs within the
volumes which today have collapsed to form the clusters. The relative
normalizations of the LFs are based on the observed relative numbers of
the different types in the field and in clusters (types 1:2:3+4 in the
proportions 0.36:0.32:032 in the field and 0.54:0.24:0.22 in clusters;
see Fig.~1) and the assumption that the total number of objects in the
initial and final LFs is the same (i.e.\ neglecting mergers within the
closed-box volume). Evolution from the initial (field) LFs to the final
(cluster) LFs occurs almost entirely through processes which suppress
star-formation as the cluster collapses and becomes denser
\citep{lew02}, converting galaxies from later types to earlier types. In
the simplest version of this model, this suppression of star-formation
does not affect the galaxies' luminosities. Fig.~8 shows the initial
(field) and final (cluster) LFs of each spectral type in this naive
model, and compares these LFs to the observed cluster LFs.

\begin{figure}
\includegraphics[width=84mm]{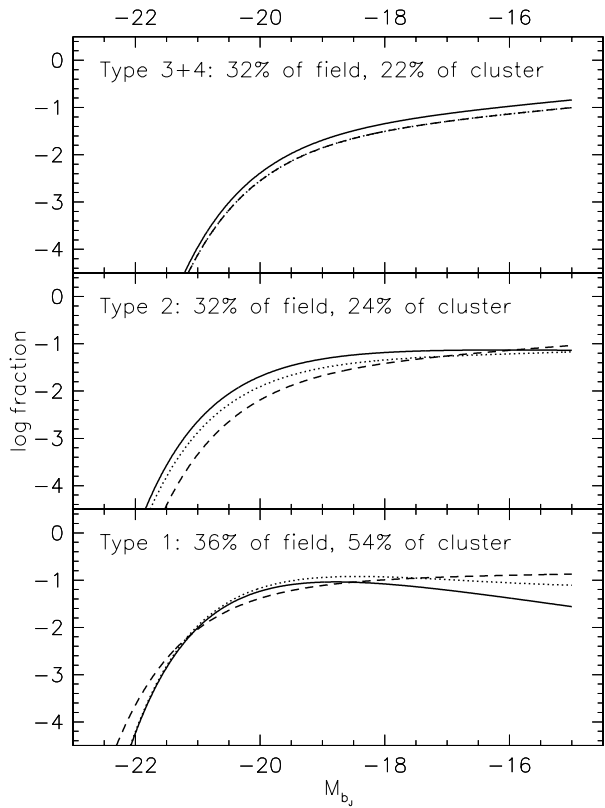}
\caption{ 
A comparison of the field LF (solid line) and closed-box model cluster
LF (dotted line) to the observed cluster LF (dashed line) for each
spectral type. Note that for type 3+4 the dotted line is identical to
the dashed line.}
\end{figure}

This model has some successes: it is consistent with the fact that the
type 3+4 cluster LF has the same shape but a lower normalization than
the field LF, and it does cause the initial field LFs of the type 1 and
2 galaxies to evolve towards the forms they are observed to have in
clusters. However this passive steepening of the LFs, achieved by
shifting galaxies from the steeper LFs of later types to the shallower
LFs of earlier types, fails to steepen the LFs sufficiently to reproduce
the observed clusters.

A more successful model requires both passive steepening as galaxies
shift from later to earlier types and also active steepening due to
luminosity-dependent fading. In such a model, the suppression of
star-formation affects about a third of type 3+4 galaxies, independent
of luminosity. The affected galaxies' instantaneous star-formation rate
decreases, lowering the strength of their $H\alpha$ emission so that
they become type 2 galaxies. This leaves the shape of the type 3+4 LF
the same, but reduces the numbers of these strongly star-forming
galaxies. When type 2 galaxies undergo further suppression of their
star-formation rate, this has the effect of decreasing their $b_{\rm J}$
luminosities and eventually converting them to type 1, where further
fading may occur. This fading must be greater for brighter galaxies in
order to actively steepen the LF slopes and reproduce the observed
cluster LFs.

In addition, some merging (or cannibalism) is required to explain the
small number of very bright type 1 galaxies in clusters which are not
present in the field LFs, and may also help explain the deficit of
bright type 2 and $L^*$ type 1 cluster galaxies. This is consistent with
the conclusions of \cite{cz02}, who find that the only difference
between the $R$-band LFs of cluster and field galaxies is an excess of
bright non-starforming galaxies. However the higher dwarf-to-giant ratio
in the LFs of earlier-type galaxies in clusters indicates that
star-formation suppression is more important than mergers in shaping the
cluster LF.

It should be noted that the field LF is in fact the mean LF of the
entire galaxy population, and is therefore dominated by galaxies
belonging to groups. Thus our results therefore imply that suppression
of star-formation is the dominant effect in evolving from the typical
group environment to the rich cluster environment. Merger effects are
expected to be more important in groups, however, and this will be
investigated in a future paper based on the group catalogue derived
directly from the 2dFGRS.

The closed-box model is over-simplified in a variety of ways, and
can only serve as a qualitative guide to understanding the processes
shaping the LF of cluster galaxies. Unfortunately, a quantitative
interpretation of our results is hampered by the fact that few
theoretical studies have considered the evolution of the LF and its
dependence on environment, and most have been limited to the brighter
cluster members (e.g.\ Malumuth \& Richstone 1984). Detailed
semi-analytic models are clearly required to explore the relative
importance of the various mechanisms that may be driving the evolution
of galaxies in clusters. These models will need to be complemented by
more stringent observational constraints to distinguish the effects of
the different processes involved. It will be particularly important to
obtain the near-UV and near-IR LFs for large spectroscopic samples of
cluster galaxies. These will yield both the instantaneous star-formation
rate and the total stellar mass of the galaxies, and reveal where
galaxies are evolving due to mergers and where they are undergoing
changes in their star-formation rate.

\section*{Acknowledgements}

We would like to thank the referee, Stefano Andreon, for a number of
suggestions which have improved the content of this paper. We also thank
Daniel Christlein and Ann Zabludoff for having shown us their results in
advance of submission. R.D.P. and W.J.C. acknowledge funding from the
Australian Research Council for this work. We are indebted to the staff
of the Anglo-Australian Observatory for their tireless efforts and
assistance in supporting 2dF throughout the course of the survey. We are
also grateful to the Australian and UK time assignment committees for
their continued support for this project.

\setlength{\bibhang}{2.0em}

\end{document}